\documentclass[apsl,showpacs,superscriptaddress,twocolumn]{revtex4-1}

\usepackage{graphicx}
\usepackage{epsfig}

\begin{document}

\title{Surface-gate-defined single-electron transistor in a MoS$_{2}$ bilayer}

\author{M. Javaid}
\email{maria.javaid@rmit.edu.au}
\affiliation{Chemical and Quantum Physics, School of Sciences, RMIT University, Melbourne VIC 3001, Australia}

\author{Daniel W. Drumm}
\affiliation{Chemical and Quantum Physics, School of Sciences, RMIT University, Melbourne VIC 3001, Australia}
\affiliation{The Australian Research Council Centre of Excellence for Nanoscale BioPhotonics, School of Sciences, RMIT University, Melbourne, VIC 3001, Australia}

\author{Salvy P. Russo}
\affiliation{Chemical and Quantum Physics, School of Sciences, RMIT University, Melbourne VIC 3001, Australia}
\author{Andrew D. Greentree}
\affiliation{Chemical and Quantum Physics, School of Sciences, RMIT University, Melbourne VIC 3001, Australia}
\affiliation{The Australian Research Council Centre of Excellence for Nanoscale BioPhotonics, School of Sciences, RMIT University, Melbourne, VIC 3001, Australia}

\date{\today}

\begin{abstract}
We report the multi-scale modeling and design of a gate-defined single-electron transistor in a MoS$_{2}$ bilayer. By combining density-functional theory and finite-element analysis, we design a surface gate structure to electrostatically define and tune a quantum dot and its associated tunnel barriers in the MoS$_{2}$ bilayer. Our approach suggests new pathways for the creation of novel quantum electronic devices in two-dimensional materials. 
\end{abstract}

\maketitle
\section{Introduction}
Novel two-dimensional (2D) materials such as graphene and transition-metal dichalcogenides (TMDCs) have attracted significant interest due to their unique electronic and optical properties \cite{Novoselov2005, Mak2010,Fiori2014}. The ultrathin geometry and dangling-bond-free interfaces of 2D materials make them good candidates to integrate on various substrates \cite{Bolotin2008}. Among these 2D materials, MoS$_{2}$ (a TMDC) is promising for quantum electronics; it possesses interesting layer-dependent properties. For example, by decreasing the thickness of MoS$_{2}$ from bulk to a single layer, its band gap switches from indirect to direct and increases by more than 0.6~eV, leading to strong photoluminescence from a single layer \cite{Mak2010, Splendiani2010}.   

The single-electron transistor (SET) is a device where electrons tunnel one by one to and from a small island through tunnel barriers \cite{Singh2012}. The tunneling of an electron and the quantization of charge are controlled by gate electrodes. SETs have extensive applications in nano-electronic devices \cite{Grabert}. They have been proposed as a future alternative to conventional CMOS transistors \cite{Mahapatra2004}, and have been used as nano-scale electrometers \cite{Devoret2000}, capable of measuring sub-electron charge variations to $10^{-6} q_{e}/\sqrt{\text {Hz}}$. They also have many applications in single-electron logic circuits \cite{Kim2012} and single-electron turnstile devices \cite{Yuan2002}. 

Many different techniques are used to fabricate SETs based in materials such as Si \cite{Fuechsle2010, Hofheinz2006, Shin2010}, GaAs \cite{Knobel2003} and carbon nanotubes \cite{Paul2008}. There is a significant body of literature on fabricating SETs. For example: Thelander \textit{et.~al.}~have designed gold-nanoparticle SETs using carbon nanotubes as leads at 200~K \cite{Thelander2001}, Klein \textit{et.~al.}~reported a colloidal chemistry technique to create cadmium-selenide nanocrystals of varying sizes \cite{Klein1997}, and Kim \textit{et.~al.}~used a focused-ion-beam technique to fabricate SETs operating at room temperature \cite{Kim2002}. Recently, there has been extensive work to define quantum-dot devices in 2D materials including graphene \cite{Stampfer2008} and TMDCs \cite{Song2015}. Surface gates have been used to create tunable tunnel barriers in Si interfaces \cite{Angus2007}, but to date there have been no reported designs of electrostatically tunable tunnel barriers and tunable quantum dots in 2D MoS$_{2}$.

It is known that the MoS$_{2}$ monolayer band structure shows no response to applied electric fields \cite{PhysRevB}. In contrast, MoS$_{2}$ bilayers exhibit significant band-structure modification under perpendicular electric field \cite{PhysRevB,Liu2012}. 

\begin{figure}[tb]
\centering
\includegraphics[width=8cm, height=11 cm]{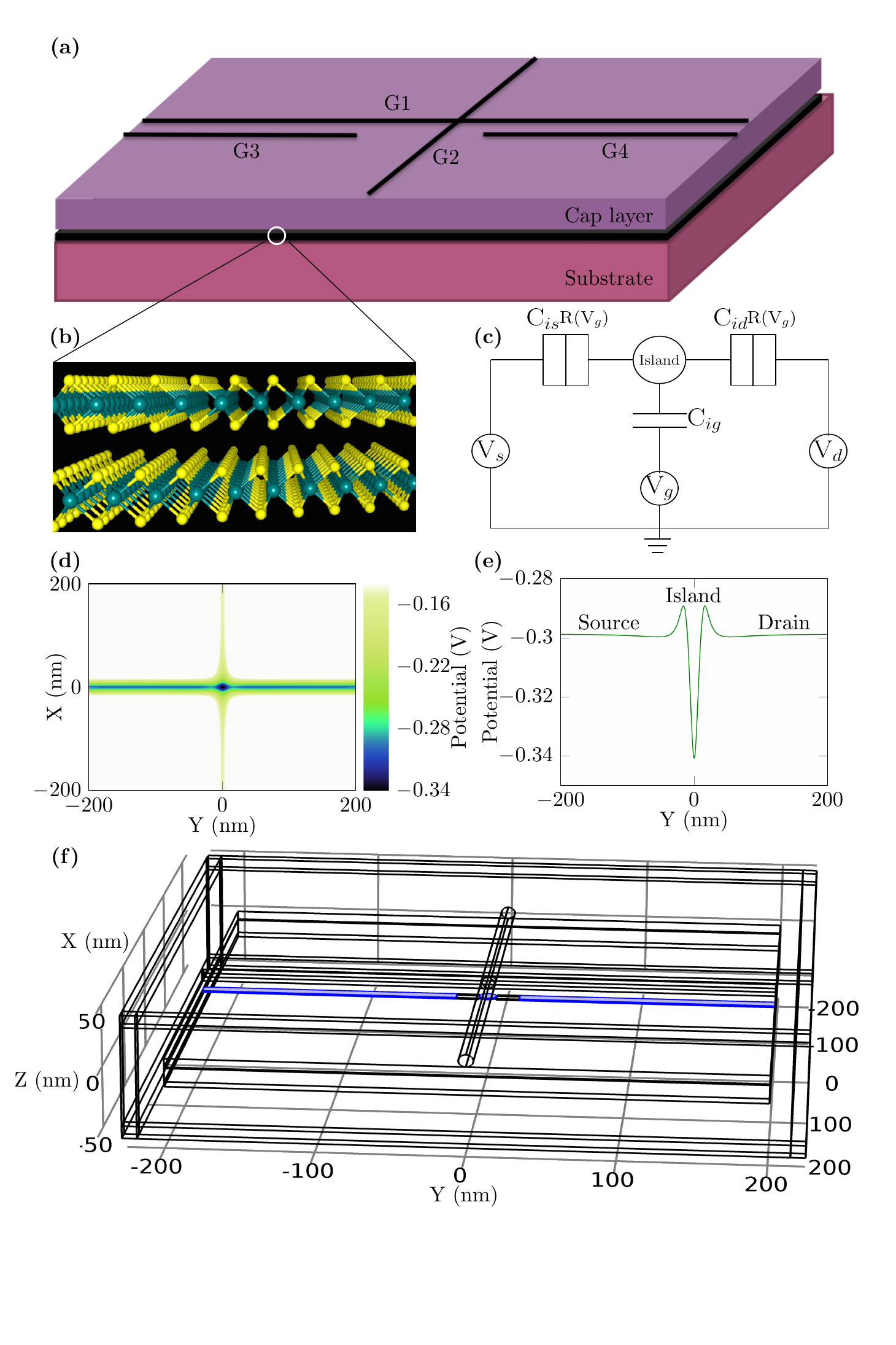} 
\caption{(Color online) (a) Schematic showing a MoS$_{2}$ bilayer sandwiched between HfO$_{2}$ cap layer and substrate, with metallic surface gates above the cap layer.  Gates G3 and G4 are shown separated from G1 for visibility, but are actually collinear.  (b) Perspective view of the black layer in (a) showing A-A$^{\prime}$ stacked, 2H MoS$_{2}$ bilayer, Mo (large, green) atoms, S (small, yellow) atoms. (c) Equivalent circuit diagram of SET showing the capacitance/resistance between the island and the source/drain and capacitance between the island and the top gates. (d) Top view of the potential created on top of the the MoS$_{2}$ bilayer sheet, showing the creation of well-defined source, drain, and island regions. (e) Potential slice along bilayer directly underneath gates G1, G3, and G4, showing source, island, and drain for a surface-gate voltage of 0.27 V. (f) 3D view of the geometry used in the \textsc{comsol} modeling. The (blue) highlighted regions are the source, island and drain in the MoS$_{2}$ bilayer plane. Lengths are in nm.}
\end{figure}

Here we present a lithographically appealing design of a surface-gate-defined SET in a MoS$_{2}$ bilayer, and model its physical characteristics. The structure of our proposed device is a MoS$_{2}$ bilayer sandwiched between a HfO$_{2}$ substrate and cap layer, with metallic surface gates, as shown in Fig.~1(a). We use the surface-gate electric fields to modify the local band structure in the MoS$_{2}$ bilayer. We present a fully self-consistent design of the device obtained through multi-scale modeling. Recently, a more conventional gate design has been used to demonstrate quantum confined structures on a few layers of WSe$_{2}$ with tunnel barriers defined by electric fields at a temperature of 240 mK \cite{Song2015}. 

We modeled a MoS$_{2}$ bilayer using density-functional theory (DFT) to study the effects of electric field on the band structure of MoS$_{2}$ bilayers. We performed numerical simulations to calculate the potential at the MoS$_{2}$ bilayer due to the surface gates and used this potential to study the modification in the lowest unoccupied molecular orbital (LUMO) of the MoS$_{2}$ bilayer. We modeled our device geometry (obtained from LUMO bending in the MoS$_{2}$ bilayer) with the commercial finite-element analysis simulation tool \textsc{comsol} \cite{comsol} to obtain the self-capacitance of the SET island and the capacitances between the island and each of the electrodes. We used these capacitances in numerical simulations of transport through the SET at a temperature of 1 K and studied its physical characteristics. We iteratively cycled through \textsc{comsol} and numerical simulations to obtain a consistent picture of the device. 

This paper is organized as follows: we begin our discussion with the DFT modeling of the MoS$_{2}$ bilayer followed by the \textsc{comsol} modeling. Then we present the influence of the potential due to surface gates in the MoS$_{2}$ bilayer plane and numerical transport simulations of the resulting SET device. 


\section{DFT Modeling of molybdenum-disulphide bilayer structures}
Although the effects of electric field on the band structure of MoS$_{2}$ bilayers are available in the literature, the focus has either been on behavior over extreme field strength ranges (with sampling too coarse for our device) \cite{PhysRevB} or on the band-gap response without specific discussion of the LUMO physics \cite{Liu2012}. Thus, to achieve the insight necessary to design our device, we explore the LUMO physics directly in the target operating regime. 

We investigated the band structure of an optimized, \mbox{A-A$^{\prime}$} stacked (Mo atoms in the top monolayer above the S atoms of the bottom monolayer), 2H-phase MoS$_{2}$ bilayer for varying electric fields. \mbox{A-A$^{\prime}$} is the most-reported stacking order for bulk MoS$_{2}$ and it is very close in energy to the most stable stacking order A-B \cite{Liu2012}. A perspective view of the MoS$_{2}$ bilayer is shown in Fig.~1 (b).
 
To model the effects of electric field on the MoS$_{2}$ bilayers, we used DFT in \textsc{crystal09} \cite{dovesi2005crystal, Dovesi2013}. MoS$_{2}$ has hexagonal symmetry and belongs to the $P6_{3}/mmc$ space group, with lattice parameters $a=3.17$~\AA \ and $c=12.324$~\AA ~\cite{Dickinson1923}. We created the bilayer unit slab by cutting a (001) plane from a bulk MoS$_{2}$ model, and including vacuum to a total cell height of $c = 500$~\AA. The exchange and correlation terms were described via the PBEsol functional \cite{Perdew2008}, which generally predicts lattice constants more accurately than PBE and LSDA (thus improving the equilibrium properties of solids), and also handles the electronic response to potentials better than most GGA functionals \cite{Perdew2008}. We used Gaussian basis sets; Mo\_SC\_HAYWSC\-311(d31)G\_cora\_1997 \cite{cora1997anab} for Mo atoms (a Hay-Wadt effective-core pseudopotential \cite{Ehlers1993} combined with a valence-electron basis set) and S\_pob\_TZVP\_2012 \cite{Peintinger2013} for S atoms (an all-electron basis set). We set a 8$\times$16$\times$1 Monkhorst-Pack \cite{Monkhorst1976} $k$-point mesh.

\begin{figure}[tb]
\centering
\includegraphics[scale=0.85]{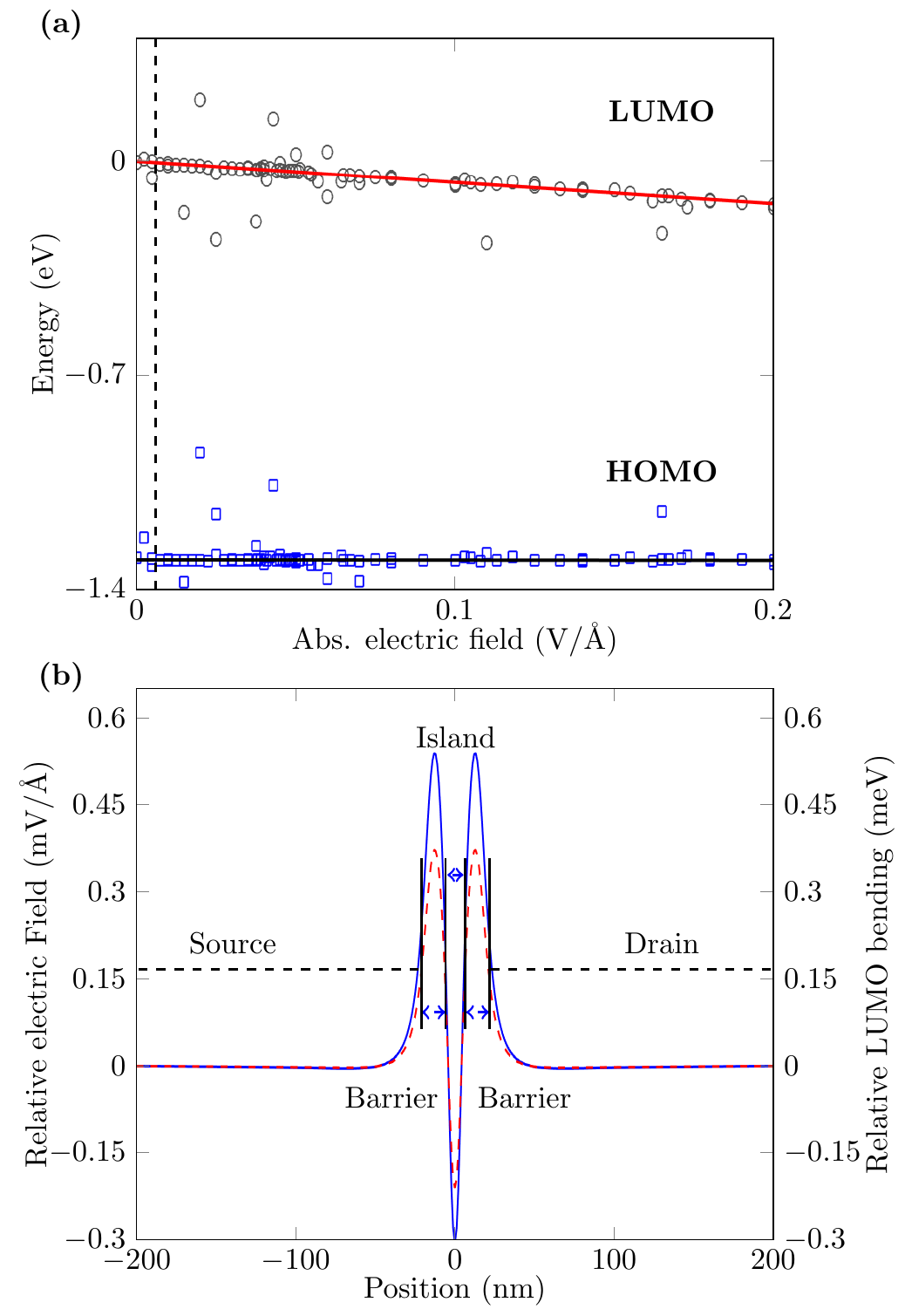}
\caption{(Color online) (a) Responses of the MoS$_{2}$ bilayer LUMO (grey circles) and HOMO (blue squares) to applied electric field, with first-order fits to LUMO (red line) and HOMO (black line) data. Negative electric-field values are folded over to the positive axis. (b) Electric field (solid blue line) relative to the dashed line in (a), created across the MoS$_{2}$ bilayer for 0.27 V surface-gate voltage. This electric field is used to calculate the LUMO bending (dashed red line) via the line-fit parameters from (a). The electric field and the LUMO bending corresponding to the intersection of dashed line with LUMO fit in (a) is set at zero. The Fermi level (black dashed line), barrier and island regions (blue dashed arrows), and region boundaries (black vertical bars) are marked.}
\end{figure}

We optimized the geometry of a MoS$_{2}$ bilayer unit slab under zero electric field and then used this optimized geometry to study the effects of electric field on the band structure of MoS$_{2}$ bilayers. We calculated the band structure along the high-symmetry path \textbf{$\Gamma$-M-K-$\Gamma$}. Band structures (shown in the Appendix) were calculated for varying electric fields applied perpendicular to the MoS$_{2}$ bilayers.

The responses of the LUMO and the highest occupied molecular orbital (HOMO) of the MoS$_{2}$ bilayer to applied electric fields are shown in Fig.~2(a). The HOMO remains static with electric fields of up to $\pm$0.2~V/\AA~(in agreement with \cite{PhysRevB}) while there is a significant response in the LUMO. Fig.~2(a) also shows some inconsistent points in the LUMO and HOMO under electric field. We excluded these points from further calculations as the band structure at these points is qualitatively different and we believe that these inconsistent points do not represent the underlying physics.

We applied a linear fit to the filtered LUMO and HOMO data, and set zero energy at the LUMO intercept (at zero field). As an electric field is applied up to $\pm$0.2~V/\AA, the LUMO bends through 138~meV. Thus electron confinement is achievable in a MoS$_{2}$ bilayer via electric-field modification of the LUMO, which occurs at a rate of 690~meV~/~(V/\AA).


\section{Circuit simulations in \textsc{comsol}}
We modeled our device geometry in \textsc{comsol} \cite{comsol} to get the capacitances between all the device components. The geometry (obtained through potential simulations discussed later and converged by iteratively cycling through numerical potential and transport simulations and \textsc{comsol}) used for \textsc{comsol} modeling is an island of diameter 12~nm and the same thickness as the MoS$_{2}$ bilayer. There are 16~nm wide MoS$_{2}$ barriers between the island and in-plane source and drain. The surface gates are modeled as two intersecting wires each of 5~nm radius sitting atop the 7~nm HfO$_{2}$ cap layer. We excluded gates G3 and G4 of Fig.~1 (a) from our \textsc{comsol} modeling as they are collinear with gate G1 and do not contribute significantly to the total capacitance of the island. The ground plane is 11~nm (thickness of the HfO$_{2}$ substrate) below the island.

The wire-frame rendering of the geometry used in \textsc{comsol} modeling is shown in Fig.~1(f), where all the lengths are in nm. 
For an island of diameter 12~nm, the self-capacitance is $C_{\rm i\Sigma}$ = 8.45~aF which corresponds to a charging energy of $e^{2}/C_{\rm i\Sigma}=18.9$~meV, where C$_{\rm i\Sigma}$ is defined as 
\begin{equation}
C_{\rm i\Sigma}= C_{\rm is} + C_{\rm id} + C_{\rm ig_{1}} + C_{\rm ig_{2}} + C_{\rm igp}
\end{equation}
 Here C$_{\rm is}$~=~$-0.252$~aF is the capacitance between the island and the buried source, C$_{\rm id}$~=~$-0.253$~aF is the capacitance between the island and the buried drain, C$_{\rm ig_{1}}$~=~$-2.2$~aF is the capacitance between the island and gate G1, C$_{\rm ig_{2}}$~=~$-2.38$~aF is the capacitance between the island and gate G2, and C$_{\rm igp}$~=~$-3.35$~aF is the capacitance between the island and the ground plane.


\section{Single-electron Transport Simulations}
In the design of our proposed MoS$_{2}$ bilayer SET, there are two longer surface gates G1 and G2, perpendicular to each other. Two shorter gates, G3 and G4, are collinear with the G1 gate as shown in Fig.~1(a). All of these surface gates are electrostatically insulated from each other; this can be done by thermally growing a thin oxide layer between them. The equivalent electrical circuit diagram is shown in Fig.~1(c), where R(V$_{g}$) is the tunnel resistance between the island and the source/drain. This resistance is a function of the gate voltage V$_{g}$ applied to the surface gates. V$_{s}$ and V$_{d}$ are the source and drain potentials applied directly to the buried source and drain in the MoS$_{2}$ bilayer plane (created by the surface gates).

We modeled our device design through in-house numerical simulations. Gates G3 and G4 were placed on top of gate G1, with their finite edges 32~nm apart from each other.  The thicknesses of the cap layer and the substrate were 7~nm and 11 nm respectively. A varying DC potential of 0.2~V to 0.34~V was applied to all surface gates. All of these parameters and the DC potential range were finalized through iterations between the \textsc{comsol} modeling and the numerical simulations. The top view of the potential created in the MoS$_{2}$ bilayer plane due to the surface gates is shown in Fig.~1(d). The intersection of gates G1 and G2 generates a well-defined island in the MoS$_{2}$ bilayer plane. Gates G1, G3, and G4 define the source/drain wiring configuration and tunable tunnel barriers in the MoS$_{2}$ bilayer plane. An analytical expression for the potential created due to one of the surface gates is given in the Appendix. 

Fig.~1(e) shows a slice of the potential on top of the MoS$_{2}$ bilayer plane and underneath gates G1, G3, and G4. Note that there is a small local minimum in the potential (and the electric field) before the barriers. The potential difference across the MoS$_{2}$ bilayer is proportional to the electric field (Fig. 2(b)). We used the linear fit to Fig. 2(a) to obtain the LUMO bending at fields shown in Fig. 2(b). This LUMO bending as a function of the position across the MoS$_{2}$ bilayer plane is shown in Fig.~2(b). With the above dimensions of the device, a 12~nm diameter island and 16~nm wide tunnel barriers are created in the MoS$_{2}$ bilayer plane. The dimensions of the SET are comparable with experimental structures in GaAs/ AlGaAs \cite{RevModPhys.75.1}. The size of the island and the width of the tunnel barriers can be controlled through the substrate and the cap layer thicknesses. The LUMO bending is used as an input to numerical transport simulation. 
 
Fig.~3(a) shows the variation of tunnel resistance with V$_{g}$. Increases to the gate voltage cause the barrier height to rise, reducing the transmission coefficient. Due to scaling of input electron energy, there is a minimal change in transmission coefficient with gate voltage but density of states in both source/drain and island increases pre-dominantly and that leads to the observed decrease in tunnel resistance. (For details of the tunnel resistance and transport calculations, see the Appendix)
\begin{figure}[tb]
\centering
\includegraphics[scale=0.9]{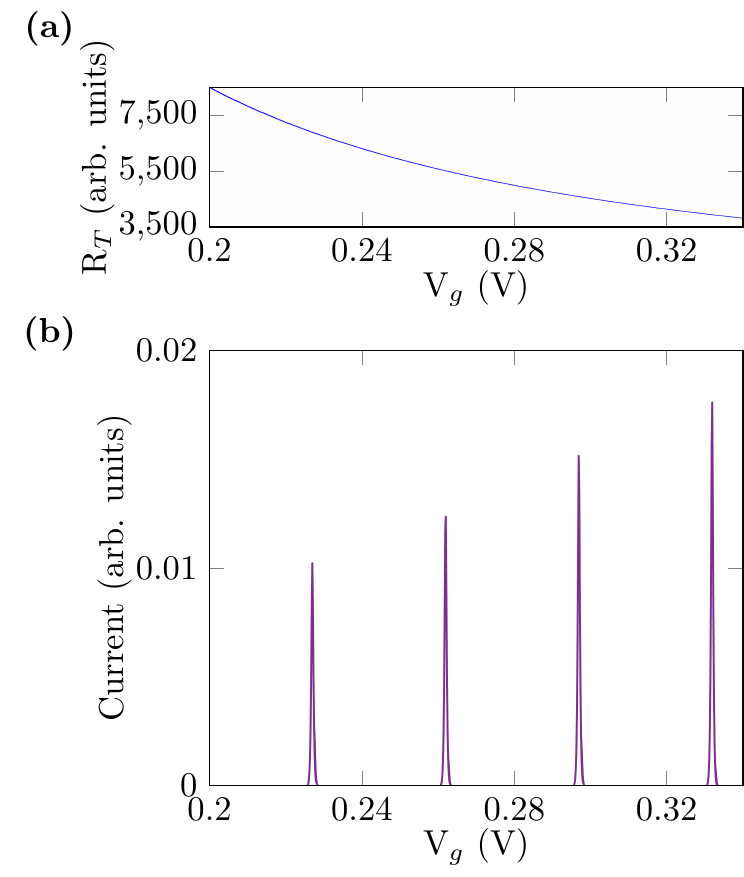}
\caption{(a) Variation of tunnel resistance with gate voltage, V$_{g}$. (b) Characteristic Coulomb blockade peaks for a bias voltage of 100~$\mu$V and at a temperature of 1~K, showing a variable current through the MoS$_{2}$ bilayer SET.}
\end{figure}
 
The behavior of the SET can be described by varying V$_{g}$. Fig.~3(b) shows the characteristic current passing through the SET at a temperature of 1~K and a bias voltage of 100~$\mu$V (In this paper, we are only considering the low-bias regime.) By varying the surface gate voltage, the overall potential landscape through the MoS$_{2}$ bilayer changes, thus modifying the barrier height and alignment between incoming electron energy and island states. When the energy of an incoming electron resonates with an unoccupied energy level in the island, that electron tunnels to the island and current flows through the SET which is shown by the peaks in Fig.~3(b). On the other hand, when none of the island states are available to the incoming electron, it cannot tunnel to the island and the current through the SET drops to zero, defining the Coulomb-blockade region between the peaks. The height variation between the current peaks arises from the changing response of the MoS$_{2}$-bilayer density of states in the source/drain and island, in contrast to an ideal metallic SET.


\section{Conclusions}

The surface gate approach to quantum devices is quite flexible and amenable to scalable integrated devices. With identical gate voltages, we already have significant control over barrier height, the energy states in the island and hence, electron tunneling. One could achieve further independent control of the incoming electron energy or island states by applying different voltages to each gate. Moreover, due to the small size of the island, our preliminary modeling holds promise for relatively high-temperature single-electron devices.  

The most important feature of the gate design shown here, as opposed to more conventional gate-defined quantum dot designs (such as, for example,  \cite{Song2015}) is that the creation of potential wells directly underneath and in the near field of the metallic surface gates should allow more-precise lithographic definition of the quantum features in the active layer.  In the conventional approach to creating sub-surface quantum dots, the relatively large distance from the gate layer to the active layer gives rise to a complicated potential landscape that must be solved for self-consistently with Schr\"{o}dinger-Poisson solvers.  Conversely, as our active layer is in the near field of the surface gates, the potential should more precisely follow the metallisation on the surface layer. This in turn means that more complicated integrated circuits should be realisable, for example multi-island dot structures or integrated qubit/readout type circuits.  This design ethos is exemplified by the surface gate defined SETs fabricated at a silicon/silica interface by Angus \textit{et al.} \cite{Angus2007}. 

Here we have studied an  embedded MoS$_{2}$ bilayer system. However, we would expect our results to be easily applicable to other bilayer two-dimensional materials that exhibit a LUMO response to electric field (\textit{e.g.} MoSe$_2$, MoTe$_2$ and WS$_2$ \cite{PhysRevB}), with only minimal modification of the applied surface potentials.


\section{Acknowledgements}
MJ and ADG acknowledge funding by ARC Discovery Grant No. D130104381. DWD acknowledges the support of the ARC Centre of Excellence for Nanoscale BioPhotonics (CE140100003). This work was supported by computational resources provided by the Australian Government through the National Computational Infrastructure under the National Computational Merit Allocation Scheme.

\appendix
\section{}

Here we present a brief overview of the design of the MoS$_{2}$ bilayer single-electron transistor, followed by details of our modeling. We show our analytic analysis of the electric field across the bilayer plane, the response of the MoS$_{2}$ bilayer band structure to electric fields, the spatial behavior of the LUMO under different gate voltages, and details of our calculations of the resistance of (and current through) the SET. 

Gates G1 and G2 of Fig. 1(a) are approximated as infinitely long lines of charge, the intersection of which defines our island via the doubly-strong steep potential. To establish source and drain leads, we use gates G3 and G4 to create similarly strong potential landscapes\\
 directly underneath gate G1.

For convenience, gates G3 and G4 are approximated as one infinite line of charge with a missing, finite segment centred on the intersection of gates G1 and G2. For our treatment, this dictates $V_{3}=V_{4}$, but that condition is not necessary if the semi-infinite gates G3 and G4 are treated separately and identical source and drain behavior is not required. It should also be noted that $V_{2}\approx V_{3}=V_{4}$ achieves comparable island and lead depths below the barrier peaks; $V_{2}=V_{3}=V_{4}$ is not a constraint. Gate G1 is therefore effectively a plunger gate controlling the depth of all components, which could in principle be operated in isolation from the rest (although $V_{1}+V_{3}>V_{2}$ is required to avoid an unintended perpendicular lead under gate G2). For simplicity here, however, we set $V_{g}=V_{1}=V_{2}=V_{3}=V_{4}$.

Defining $x$ as the distance in the bilayer plane transverse to gate G1 (with zero value directly underneath G1), ($y$ similarly with reference to G2), and $z$ as the bilayer-ground plane separation, the electric potential difference created across the MoS$_{2}$ bilayers is the sum effect from all gates, which is analytic. For example, the potential due to gate G1 is:
\begin{equation}
V=\frac{\lambda}{2\pi\epsilon_{o}\epsilon_{r}}\text{log}\left[\frac{x^{2}+(z-d)^{2}}{x^{2}+(z+d)^{2}}\right],
\end{equation}
where $\lambda$ is the linear charge density of gate G1, $\epsilon_{r}$ is the dielectric constant of HfO$_{2}$ \cite{Jones2005} and $d $ is the distance of gate G1 from the ground plane.  The other potentials are similarly derived, and the full potential is the sum effect from all of the gates. The finite difference between the top and bottom edges of the bilayer is evaluated to provide the field strength across the bilayer.

\begin{figure}[tb!]
\centering
\includegraphics[width=0.99\linewidth]{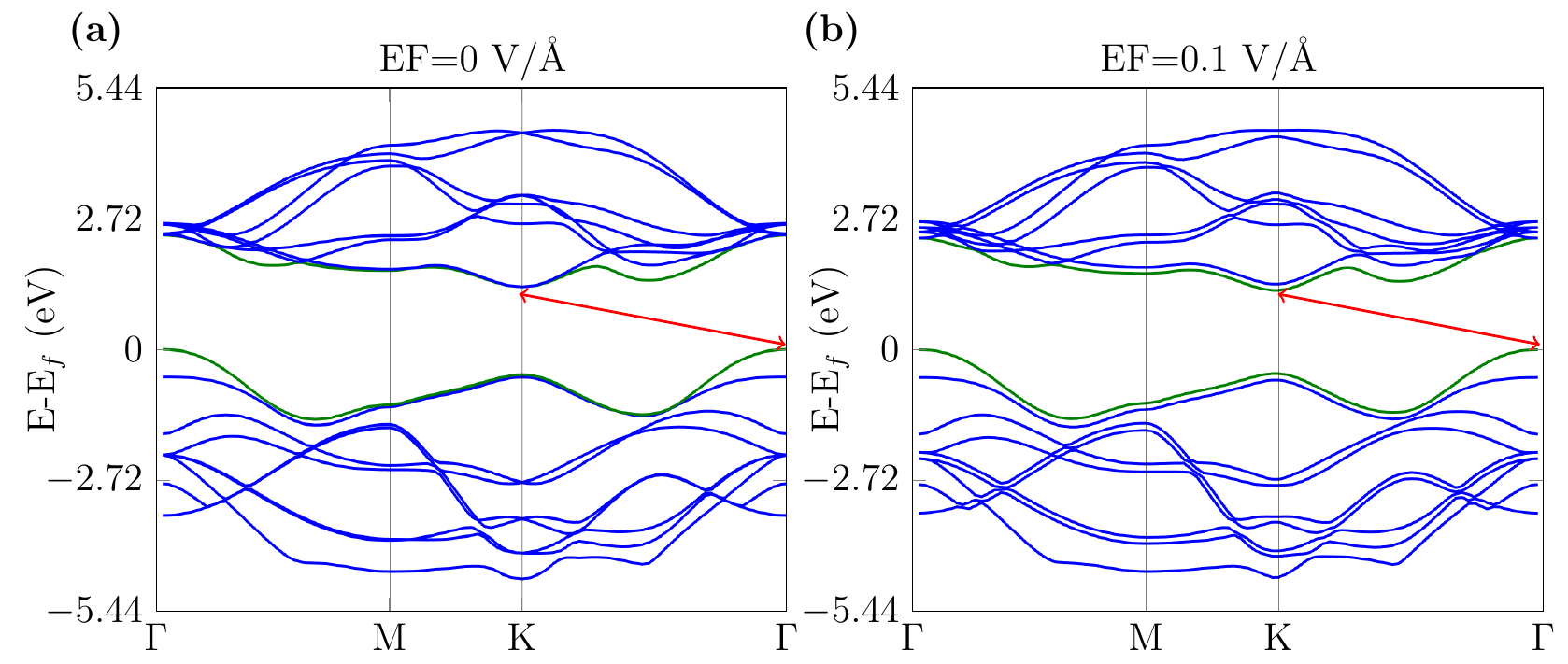}
\caption{(a) Band structure of MoS$_{2}$ bilayer under zero electric field. There is an indirect band gap of 1.30 eV. (b) Band structure of MoS$_{2}$ bilayer under finite electric field. There is an indirect band gap of 1.23 eV.}
\label{fig:FigS1}
\end{figure}

Having obtained a two-dimensional map of the potential difference across the bilayer plane -- \textit{e.g.}, Fig. 1(d) in the main paper -- we then consider its effect on the MoS$_{2}$ band structure at fields from 0 to 0.2~V/\r{A} in steps no larger than 0.01~V/\r{A}, and smaller as required to verify linear behavior.  Fig. \ref{fig:FigS1} shows two sample band structures.

We are particularly interested in the minimal substrate footprint, and the constraint of reasonable applied field strength. Therefore in the main paper, without loss of generality in the method, we explicitly consider the high charging energy case, when the classical occupancy of the SET island can change by at most one electron at a time. Thus, the extent of the LUMO bending created must match well to the charging energy of the system;
\begin{equation}
\Delta L\approx E=\frac{1}{2}Q^{T}C^{-1}Q,
\end{equation}
where $Q$ is the charge on the SET, defined by $Q$=[$Q_{i}$~$Q_{s}$~$Q_{g1}$~$Q_{g2}$~$Q_{d}$~$Q_{gp}$]$^{T}$. $C$ is the capacitance matrix between all the device components and is defined as:
\begin{equation}
C= \left( \begin{array}{cccccc}
C_{\rm i\Sigma} & C_{\rm is} & C_{\rm ig1} & C_{\rm ig2} & C_{\rm id} &  C_{\rm igp}\\
C_{\rm si} & C_{\rm s\Sigma} &  C_{\rm sg1} & C_{\rm sg2} & C_{\rm sd} &  C_{\rm sgp}\\
C_{\rm g1i} & C_{\rm g1s} &  C_{\rm g1\Sigma} & C_{\rm g1g2} & C_{\rm g1d} &  C_{\rm g1gp}\\
C_{\rm g2i} & C_{\rm g2s} &  C_{\rm g2g1} & C_{\rm g2\Sigma} & C_{\rm g2d} &  C_{\rm g2gp} \\
C_{\rm di} & C_{\rm ds} &  C_{\rm dg1} & C_{\rm dg2} & C_{\rm d\Sigma} &  C_{\rm dgp}\\
C_{\rm gpi} & C_{\rm gps} &  C_{\rm gpg1} & C_{\rm gpg2} & C_{\rm gpd} &  C_{\rm gp\Sigma}\end{array} \right)
\end{equation}
the terms of which are obtained via \textsc{comsol} modeling as described in the main paper.

The tunnel resistance cannot be evaluated exactly. The usual approach to an arbitrary barrier, the WKB approximation, is not directly applicable if the incoming energy of the electron matches the barrier energy at any point. In fact, not only does the approximation not hold at those points, but also in their vicinities. The connection formulas, using an Airy function ansatz to patch difficult regions, are the standard way to circumvent this issue; however, the conditions for their use are not met by this barrier. Accordingly, we make the approximation that the barrier is finite and rectangular, with the identical width (16~nm) and height equal to the peak value of LUMO bending, called $L_{{\rm peak}}$ while setting $L(-200)$~=~0. Since such a barrier is definitively larger, and no thinner, than the original barrier, its transmission coefficient is guaranteed to be lower than any experimental device. The transmission coefficient of a finite square barrier is:
\begin{equation}
T = e^{-2\gamma},
\end{equation}
where 
\begin{equation}
\gamma=\frac{1}{\hbar}\int\limits_{0}^{a}\sqrt{2m(L_{{\rm peak}}-\varepsilon)}dx=\frac{a}{\hbar}\sqrt{2m\left(L_{{\rm peak}}-\varepsilon\right)},
\end{equation}
 $\hbar=h/2\pi$ is the reduced Planck's constant, $m= 0.38m_{o}$ is the effective mass of electrons in the MoS$_{2}$ bilayer \cite{mass}, $a$ is the width of the tunnel barrier, and $\varepsilon$ is the energy of an incoming electron in the MoS$_{2}$ bilayer.
 
\begin{figure}[tb!]
\centering
\includegraphics[width=0.99\linewidth]{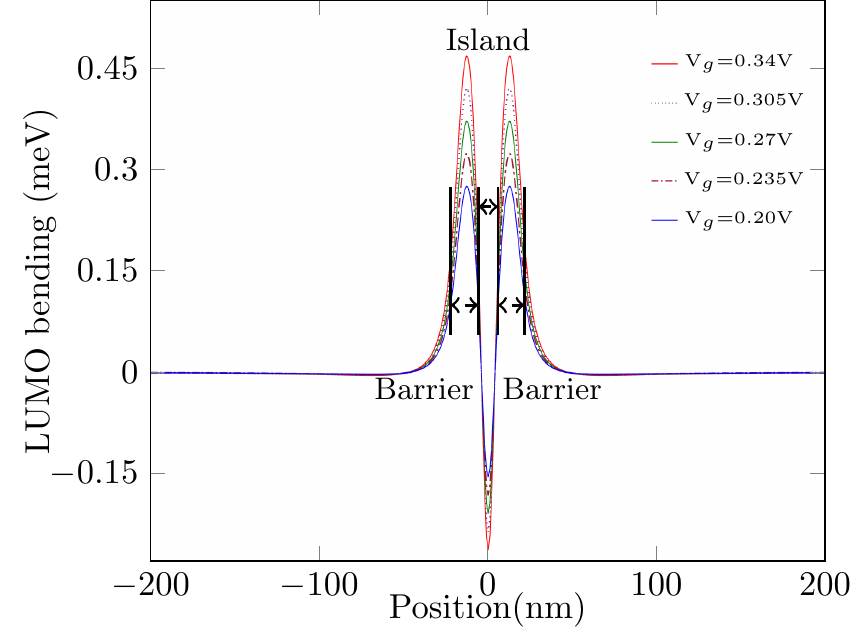}
\caption{(Color online) LUMO bending as function of position along the source-island-drain axis for varying gate voltages within the device target operating regime. The barrier function behaves linearly with $V_{g}$. Vertical marks show the boundaries of the labelled regions (black-dashed lines with arrow heads).  $\varepsilon$ is not shown since it varies with $V_{g}$, but its values for each $V_{g}$ can be identified as the intersections of the corresponding LUMO data traces with the region boundaries.}
\label{fig:S2}
\end{figure}

 Here $\varepsilon$ can also be regarded as $\varepsilon=V_{sd}+E_{{\rm th}} + \mu$, where $V_{sd}$ is the source-drain bias, $E_{{\rm th}}$ is the thermal energy described by the Fermi smearing incorporated below, and $\mu$ is the chemical potential of the electron which is set by the absolute source and drain voltages $V_{s}$ and $V_{d}$. (Recall $V_{sd}=V_{s}-V_{d}$; modifying the absolute value of each in tandem has no effect on $V_{sd}$.) We assume that the incoming electron has $\mu$ such that $\varepsilon\approx0.38\left[L_{{\rm peak}}-L\left(-200\right)\right]$~=~0.38$L_{{\rm peak}}$; this defines the points where $L=\varepsilon$ (the region boundaries) at consistent locations in space -- enforcing the desired geometry with large charging energy. (The left edge of $L$ has been set as zero energy, the absolute value of $\mu$ is somewhat different to the marked $\varepsilon$.) Fig. \ref{fig:S2} shows $L\left(x\right)$ for several values of $V_{g}$, together with the constant region boundaries. In principle, this assumption could be relaxed as needed to describe a particular physical device.

As an example, Fig. 2(b) in the main text shows the particular case of the optimised structure under $V_{g}$~=~$0.27$~V, where $\mu$ is set such that $\varepsilon$ is equal to the black-dashed line.  There, the island has 12~nm diameter and 16~nm wide tunnel barriers. $L$ is the barrier height created by LUMO bending due to surface gate potentials. 

The tunnel resistance is given by \cite{thesis}
\begin{equation}
\label{eq:resistance}
R_{T}^{-1}=\frac{2\pi e^{2}}{\hbar}\rho_{1}\rho_{2}\mid T(\varepsilon)\mid^{2},
\end{equation}
where $\mid T(\varepsilon)\mid$ is the transmission co-efficient, $\rho_{1}$ is the density of states in source/drain electrodes and $\rho_{2}$ is the density of states in the island. We approximate the density of states by assuming constant band bending over the leads (and island), giving:
\begin{equation}
\label{eq:density}
\rho=\int \limits_{\varepsilon_{o}}^{\varepsilon_{f}} D(\varepsilon)~d\varepsilon = \frac{2}{3}D(\varepsilon-\varepsilon_{o})\times (\varepsilon-\varepsilon_{o});
\end{equation}
Here $\varepsilon_{f}$ is the Fermi level and $\varepsilon_{o}$ is the energy of the lowest level in the source/drain leads of MoS$_{2}$ bilayer plane. (This corresponds to making the assumption that electron-hole recombination is negligibly slow compared to the tunneling rate). $D(\varepsilon)$ is the density of states per unit energy in a solid, given by
\begin{equation}
\label{eq:DOS}
D(\varepsilon) = \frac{8\pi\sqrt{2}}{h^{3}}m^{\frac{3}{2}}\sqrt{\varepsilon}\times \eta \text{ for } \varepsilon>\varepsilon_{o}, 
\end{equation}

where $\eta$ is the volume.  

Hence,
\begin{equation}
\rho_{1}= \frac{16\pi\sqrt{2}}{3h^{3}}m^{\frac{3}{2}} \sqrt{\varepsilon-\varepsilon_{o}}(\varepsilon-\varepsilon_{o})\times \eta_{s} \text{ for } \varepsilon>\varepsilon_{o}
\end{equation}
and
\begin{equation}
\rho_{2}= \frac{16\pi\sqrt{2}}{3h^{3}}m^{\frac{3}{2}} \sqrt{\varepsilon-\varepsilon_{o}}(\varepsilon-\varepsilon_{o})\times \eta_{i} \text{ for } \varepsilon>\varepsilon_{o}
\end{equation}
Here $\eta_{s}$ and $\eta_{i}$ are the volumes for the source and island respectively. Volumes were estimated assuming the bilayer height of 0.94~\r{A}, and island and lead in-plane areas of $\pi\left(6{\rm ~nm}\right)^{2}$ and $400\times 20{\rm ~nm}^{2}$, respectively which leads to tunnel resistance of 3500~M$\Omega$ to 8500~M$\Omega$ for this idealised design in perfect MoS$_{2}$ in HfO$_{2}$. Practically, many factors involved in experiment (\textit{e.g.}, fabrication imprecision, defects, \textit{etc.}) often accumulate to affect resistances by several orders of magnitude. 

Although we use a semiconductor material (MoS$_{2}$ bilayer) in the design of our SET, after connecting it with the external batteries the Fermi level rises into the conduction band in the source/island/drain regions while sitting below the conduction band in the tunnel barrier regions.  Thus, the SET behaves effectively like a metal-insulator-metal junction and all the standard approximations for metallic SET are still applicable to our SET. 
 
To determine the current through the SET, we need the tunneling rates on and off the island for different charge configurations of the SET. These tunneling rates are given by \cite{Grabert}
\begin{equation}
\label{eq:rate}
\Gamma_{\rm \chi i}^{n}=\frac{1}{q_{e}^{2}R_{T}}\frac{\Delta E_{\rm \chi i}^{n}}{\text{exp}(\frac{\Delta E_{\rm \chi i}^{n}}{k_{B}T})-1},
\end{equation}
where $\chi$ denotes the source or drain and i the island. $\Delta E_{\rm \chi i}^{n}$ is the energy difference between different charge configurations of the SET and is given by
\begin{equation}
\Delta E_{\rm \chi i}^{n}=E(n-1)-E(n) + V_{\rm \chi}q_{e}
\end{equation} 
Here $V_{\chi}$ is the chemical potential applied to the source/drain lead.
  
The current through the SET is sum of the probabilities for all possible tunneling rates on and off the island. This current is given by
\begin{equation}
I=q_{e}\sum_{\rm n=-\infty}^{\infty}\sum_{\rm \chi=s,d}p_{n}(\Gamma_{\rm \chi i}^{n}-\Gamma_{\rm i\chi}^{n})
\end{equation}
where $p_{n}$ is the probability that island is in a state with $n$ excess electrons. These probabilities are calculated by the master equation as described in \cite{Conrad05}.

\bibliography{SETref}

\begin{thebibliography}{38}%
\makeatletter
\providecommand \@ifxundefined [1]{%
 \@ifx{#1\undefined}
}%
\providecommand \@ifnum [1]{%
 \ifnum #1\expandafter \@firstoftwo
 \else \expandafter \@secondoftwo
 \fi
}%
\providecommand \@ifx [1]{%
 \ifx #1\expandafter \@firstoftwo
 \else \expandafter \@secondoftwo
 \fi
}%
\providecommand \natexlab [1]{#1}%
\providecommand \enquote  [1]{``#1''}%
\providecommand \bibnamefont  [1]{#1}%
\providecommand \bibfnamefont [1]{#1}%
\providecommand \citenamefont [1]{#1}%
\providecommand \href@noop [0]{\@secondoftwo}%
\providecommand \href [0]{\begingroup \@sanitize@url \@href}%
\providecommand \@href[1]{\@@startlink{#1}\@@href}%
\providecommand \@@href[1]{\endgroup#1\@@endlink}%
\providecommand \@sanitize@url [0]{\catcode `\\12\catcode `\$12\catcode
  `\&12\catcode `\#12\catcode `\^12\catcode `\_12\catcode `\%12\relax}%
\providecommand \@@startlink[1]{}%
\providecommand \@@endlink[0]{}%
\providecommand \url  [0]{\begingroup\@sanitize@url \@url }%
\providecommand \@url [1]{\endgroup\@href {#1}{\urlprefix }}%
\providecommand \urlprefix  [0]{URL }%
\providecommand \Eprint [0]{\href }%
\providecommand \doibase [0]{http://dx.doi.org/}%
\providecommand \selectlanguage [0]{\@gobble}%
\providecommand \bibinfo  [0]{\@secondoftwo}%
\providecommand \bibfield  [0]{\@secondoftwo}%
\providecommand \translation [1]{[#1]}%
\providecommand \BibitemOpen [0]{}%
\providecommand \bibitemStop [0]{}%
\providecommand \bibitemNoStop [0]{.\EOS\space}%
\providecommand \EOS [0]{\spacefactor3000\relax}%
\providecommand \BibitemShut  [1]{\csname bibitem#1\endcsname}%
\let\auto@bib@innerbib\@empty
\bibitem [{\citenamefont {Novoselov}\ \emph {et~al.}(2005)\citenamefont
  {Novoselov}, \citenamefont {Geim}, \citenamefont {Morozov}, \citenamefont
  {Jiang}, \citenamefont {Katsnelson}, \citenamefont {Grigorieva},
  \citenamefont {Dubonos},\ and\ \citenamefont {Firsov}}]{Novoselov2005}%
  \BibitemOpen
  \bibfield  {author} {\bibinfo {author} {\bibfnamefont {K.~S.}\ \bibnamefont
  {Novoselov}}, \bibinfo {author} {\bibfnamefont {A.~K.}\ \bibnamefont {Geim}},
  \bibinfo {author} {\bibfnamefont {S.~V.}\ \bibnamefont {Morozov}}, \bibinfo
  {author} {\bibfnamefont {D.}~\bibnamefont {Jiang}}, \bibinfo {author}
  {\bibfnamefont {M.~I.}\ \bibnamefont {Katsnelson}}, \bibinfo {author}
  {\bibfnamefont {I.~V.}\ \bibnamefont {Grigorieva}}, \bibinfo {author}
  {\bibfnamefont {S.~V.}\ \bibnamefont {Dubonos}}, \ and\ \bibinfo {author}
  {\bibfnamefont {A.~A.}\ \bibnamefont {Firsov}},\ }\href
  {http://dx.doi.org/10.1038/nature04233} {\bibfield  {journal} {\bibinfo
  {journal} {Nature}\ }\textbf {\bibinfo {volume} {438}},\ \bibinfo {pages}
  {197} (\bibinfo {year} {2005})}\BibitemShut {NoStop}%
\bibitem [{\citenamefont {Mak}\ \emph {et~al.}(2010)\citenamefont {Mak},
  \citenamefont {Lee}, \citenamefont {Hone}, \citenamefont {Shan},\ and\
  \citenamefont {Heinz}}]{Mak2010}%
  \BibitemOpen
  \bibfield  {author} {\bibinfo {author} {\bibfnamefont {K.~F.}\ \bibnamefont
  {Mak}}, \bibinfo {author} {\bibfnamefont {C.}~\bibnamefont {Lee}}, \bibinfo
  {author} {\bibfnamefont {J.}~\bibnamefont {Hone}}, \bibinfo {author}
  {\bibfnamefont {J.}~\bibnamefont {Shan}}, \ and\ \bibinfo {author}
  {\bibfnamefont {T.~F.}\ \bibnamefont {Heinz}},\ }\href {\doibase
  10.1103/PhysRevLett.105.136805} {\bibfield  {journal} {\bibinfo  {journal}
  {Phy. Rev. Lett.}\ }\textbf {\bibinfo {volume} {105}},\ \bibinfo {pages}
  {136805} (\bibinfo {year} {2010})}\BibitemShut {NoStop}%
\bibitem [{\citenamefont {Fiori}\ \emph {et~al.}(2014)\citenamefont {Fiori},
  \citenamefont {Bonaccorso}, \citenamefont {Iannaccone}, \citenamefont
  {Palacios}, \citenamefont {Neumaier}, \citenamefont {Seabaugh}, \citenamefont
  {Banerjee},\ and\ \citenamefont {Colombo}}]{Fiori2014}%
  \BibitemOpen
  \bibfield  {author} {\bibinfo {author} {\bibfnamefont {G.}~\bibnamefont
  {Fiori}}, \bibinfo {author} {\bibfnamefont {F.}~\bibnamefont {Bonaccorso}},
  \bibinfo {author} {\bibfnamefont {G.}~\bibnamefont {Iannaccone}}, \bibinfo
  {author} {\bibfnamefont {T.}~\bibnamefont {Palacios}}, \bibinfo {author}
  {\bibfnamefont {D.}~\bibnamefont {Neumaier}}, \bibinfo {author}
  {\bibfnamefont {A.}~\bibnamefont {Seabaugh}}, \bibinfo {author}
  {\bibfnamefont {S.~K.}\ \bibnamefont {Banerjee}}, \ and\ \bibinfo {author}
  {\bibfnamefont {L.}~\bibnamefont {Colombo}},\ }\href {\doibase
  10.1038/nnano.2014.207} {\bibfield  {journal} {\bibinfo  {journal} {Nat.
  Nanotechnol.}\ }\textbf {\bibinfo {volume} {9}},\ \bibinfo {pages} {768}
  (\bibinfo {year} {2014})}\BibitemShut {NoStop}%
\bibitem [{\citenamefont {Bolotin}\ \emph {et~al.}(2008)\citenamefont
  {Bolotin}, \citenamefont {Sikes}, \citenamefont {Jiang}, \citenamefont
  {Klima}, \citenamefont {Fudenberg}, \citenamefont {Hone}, \citenamefont
  {Kim},\ and\ \citenamefont {Stormer}}]{Bolotin2008}%
  \BibitemOpen
  \bibfield  {author} {\bibinfo {author} {\bibfnamefont {K.}~\bibnamefont
  {Bolotin}}, \bibinfo {author} {\bibfnamefont {K.}~\bibnamefont {Sikes}},
  \bibinfo {author} {\bibfnamefont {Z.}~\bibnamefont {Jiang}}, \bibinfo
  {author} {\bibfnamefont {M.}~\bibnamefont {Klima}}, \bibinfo {author}
  {\bibfnamefont {G.}~\bibnamefont {Fudenberg}}, \bibinfo {author}
  {\bibfnamefont {J.}~\bibnamefont {Hone}}, \bibinfo {author} {\bibfnamefont
  {P.}~\bibnamefont {Kim}}, \ and\ \bibinfo {author} {\bibfnamefont
  {H.}~\bibnamefont {Stormer}},\ }\href {\doibase 10.1016/j.ssc.2008.02.024}
  {\bibfield  {journal} {\bibinfo  {journal} {Solid State Commun.}\ }\textbf
  {\bibinfo {volume} {146}},\ \bibinfo {pages} {351} (\bibinfo {year}
  {2008})}\BibitemShut {NoStop}%
\bibitem [{\citenamefont {Splendiani}\ \emph {et~al.}(2010)\citenamefont
  {Splendiani}, \citenamefont {Sun}, \citenamefont {Zhang}, \citenamefont {Li},
  \citenamefont {Kim}, \citenamefont {Chim}, \citenamefont {Galli},\ and\
  \citenamefont {Wang}}]{Splendiani2010}%
  \BibitemOpen
  \bibfield  {author} {\bibinfo {author} {\bibfnamefont {A.}~\bibnamefont
  {Splendiani}}, \bibinfo {author} {\bibfnamefont {L.}~\bibnamefont {Sun}},
  \bibinfo {author} {\bibfnamefont {Y.}~\bibnamefont {Zhang}}, \bibinfo
  {author} {\bibfnamefont {T.}~\bibnamefont {Li}}, \bibinfo {author}
  {\bibfnamefont {J.}~\bibnamefont {Kim}}, \bibinfo {author} {\bibfnamefont
  {C.~Y.}\ \bibnamefont {Chim}}, \bibinfo {author} {\bibfnamefont
  {G.}~\bibnamefont {Galli}}, \ and\ \bibinfo {author} {\bibfnamefont
  {F.}~\bibnamefont {Wang}},\ }\href {\doibase 10.1021/nl903868w} {\bibfield
  {journal} {\bibinfo  {journal} {Nano Lett.}\ }\textbf {\bibinfo {volume}
  {10}},\ \bibinfo {pages} {1271} (\bibinfo {year} {2010})}\BibitemShut
  {NoStop}%
\bibitem [{\citenamefont {Singh}\ \emph {et~al.}(2012)\citenamefont {Singh},
  \citenamefont {Agrawal},\ and\ \citenamefont {Singh}}]{Singh2012}%
  \BibitemOpen
  \bibfield  {author} {\bibinfo {author} {\bibfnamefont {V.}~\bibnamefont
  {Singh}}, \bibinfo {author} {\bibfnamefont {A.}~\bibnamefont {Agrawal}}, \
  and\ \bibinfo {author} {\bibfnamefont {S.}~\bibnamefont {Singh}},\ }\href
  {http://www.ijsce.org/attachments/File/v2i3/C0831062312.pdf} {\bibfield
  {journal} {\bibinfo  {journal} {IJSCE}\ }\textbf {\bibinfo {volume} {2}},\
  \bibinfo {pages} {502} (\bibinfo {year} {2012})}\BibitemShut {NoStop}%
\bibitem [{\citenamefont {Grabert}\ and\ \citenamefont
  {Devoret}(1992)}]{Grabert}%
  \BibitemOpen
  \bibfield  {author} {\bibinfo {author} {\bibfnamefont {H.}~\bibnamefont
  {Grabert}}\ and\ \bibinfo {author} {\bibfnamefont {M.~H.}\ \bibnamefont
  {Devoret}},\ }\href@noop {} {}\ (\bibinfo  {publisher} {Plenum Press, New
  York},\ \bibinfo {year} {1992})\BibitemShut {NoStop}%
\bibitem [{\citenamefont {Mahapatra}\ \emph {et~al.}(2004)\citenamefont
  {Mahapatra}, \citenamefont {Vaish}, \citenamefont {Wasshuber}, \citenamefont
  {Banerjee},\ and\ \citenamefont {Ionescu}}]{Mahapatra2004}%
  \BibitemOpen
  \bibfield  {author} {\bibinfo {author} {\bibfnamefont {S.}~\bibnamefont
  {Mahapatra}}, \bibinfo {author} {\bibfnamefont {V.}~\bibnamefont {Vaish}},
  \bibinfo {author} {\bibfnamefont {C.}~\bibnamefont {Wasshuber}}, \bibinfo
  {author} {\bibfnamefont {K.}~\bibnamefont {Banerjee}}, \ and\ \bibinfo
  {author} {\bibfnamefont {A.~M.}\ \bibnamefont {Ionescu}},\ }\href {\doibase
  10.1109/TED.2004.837369} {\bibfield  {journal} {\bibinfo  {journal} {IEEE
  Trans. Electron Devices}\ }\textbf {\bibinfo {volume} {51}},\ \bibinfo
  {pages} {1772} (\bibinfo {year} {2004})}\BibitemShut {NoStop}%
\bibitem [{\citenamefont {Devoret}\ and\ \citenamefont
  {Schoelkopf}(2000)}]{Devoret2000}%
  \BibitemOpen
  \bibfield  {author} {\bibinfo {author} {\bibfnamefont {M.}~\bibnamefont
  {Devoret}}\ and\ \bibinfo {author} {\bibfnamefont {R.}~\bibnamefont
  {Schoelkopf}},\ }\href {\doibase 10.1038/35023253} {\bibfield  {journal}
  {\bibinfo  {journal} {Nature}\ }\textbf {\bibinfo {volume} {406}},\ \bibinfo
  {pages} {1039} (\bibinfo {year} {2000})}\BibitemShut {NoStop}%
\bibitem [{\citenamefont {Kim}\ \emph {et~al.}(2012)\citenamefont {Kim},
  \citenamefont {Lee}, \citenamefont {Kang}, \citenamefont {Choi},
  \citenamefont {Yu}, \citenamefont {Takahashi},\ and\ \citenamefont
  {Hasko}}]{Kim2012}%
  \BibitemOpen
  \bibfield  {author} {\bibinfo {author} {\bibfnamefont {S.~J.}\ \bibnamefont
  {Kim}}, \bibinfo {author} {\bibfnamefont {J.~J.}\ \bibnamefont {Lee}},
  \bibinfo {author} {\bibfnamefont {H.~J.}\ \bibnamefont {Kang}}, \bibinfo
  {author} {\bibfnamefont {J.~B.}\ \bibnamefont {Choi}}, \bibinfo {author}
  {\bibfnamefont {Y.-S.}\ \bibnamefont {Yu}}, \bibinfo {author} {\bibfnamefont
  {Y.}~\bibnamefont {Takahashi}}, \ and\ \bibinfo {author} {\bibfnamefont
  {D.~G.}\ \bibnamefont {Hasko}},\ }\href {\doibase 10.1063/1.4761935}
  {\bibfield  {journal} {\bibinfo  {journal} {Appl. Phys. Lett.}\ }\textbf
  {\bibinfo {volume} {101}},\ \bibinfo {pages} {183101} (\bibinfo {year}
  {2012})}\BibitemShut {NoStop}%
\bibitem [{\citenamefont {Yuan}(2002)}]{Yuan2002}%
  \BibitemOpen
  \bibfield  {author} {\bibinfo {author} {\bibfnamefont {Z.~L.}\ \bibnamefont
  {Yuan}},\ }\href {\doibase 10.1126/science.1066790} {\bibfield  {journal}
  {\bibinfo  {journal} {Science}\ }\textbf {\bibinfo {volume} {295}},\ \bibinfo
  {pages} {102} (\bibinfo {year} {2002})}\BibitemShut {NoStop}%
\bibitem [{\citenamefont {Fuechsle}\ \emph {et~al.}(2010)\citenamefont
  {Fuechsle}, \citenamefont {Mahapatra}, \citenamefont {Zwanenburg},
  \citenamefont {Friesen}, \citenamefont {Eriksson},\ and\ \citenamefont
  {Simmons}}]{Fuechsle2010}%
  \BibitemOpen
  \bibfield  {author} {\bibinfo {author} {\bibfnamefont {M.}~\bibnamefont
  {Fuechsle}}, \bibinfo {author} {\bibfnamefont {S.}~\bibnamefont {Mahapatra}},
  \bibinfo {author} {\bibfnamefont {F.~A.}\ \bibnamefont {Zwanenburg}},
  \bibinfo {author} {\bibfnamefont {M.}~\bibnamefont {Friesen}}, \bibinfo
  {author} {\bibfnamefont {M.~A.}\ \bibnamefont {Eriksson}}, \ and\ \bibinfo
  {author} {\bibfnamefont {M.~Y.}\ \bibnamefont {Simmons}},\ }\href {\doibase
  10.1038/nnano.2010.95} {\bibfield  {journal} {\bibinfo  {journal} {Nat.
  Nanotechnol.}\ }\textbf {\bibinfo {volume} {5}},\ \bibinfo {pages} {502}
  (\bibinfo {year} {2010})}\BibitemShut {NoStop}%
\bibitem [{\citenamefont {Hofheinz}\ \emph {et~al.}(2006)\citenamefont
  {Hofheinz}, \citenamefont {Jehl}, \citenamefont {Sanquer}, \citenamefont
  {Molas}, \citenamefont {Vinet},\ and\ \citenamefont
  {Deleonibus}}]{Hofheinz2006}%
  \BibitemOpen
  \bibfield  {author} {\bibinfo {author} {\bibfnamefont {M.}~\bibnamefont
  {Hofheinz}}, \bibinfo {author} {\bibfnamefont {X.}~\bibnamefont {Jehl}},
  \bibinfo {author} {\bibfnamefont {M.}~\bibnamefont {Sanquer}}, \bibinfo
  {author} {\bibfnamefont {G.}~\bibnamefont {Molas}}, \bibinfo {author}
  {\bibfnamefont {M.}~\bibnamefont {Vinet}}, \ and\ \bibinfo {author}
  {\bibfnamefont {S.}~\bibnamefont {Deleonibus}},\ }\href
  {http://scitation.aip.org/content/aip/journal/apl/89/14/10.1063/1.2358812}
  {\bibfield  {journal} {\bibinfo  {journal} {Appl. Phys. Lett.}\ }\textbf
  {\bibinfo {volume} {89}},\ \bibinfo {eid} {143504} (\bibinfo {year}
  {2006})}\BibitemShut {NoStop}%
\bibitem [{\citenamefont {Shin}\ \emph {et~al.}(2010)\citenamefont {Shin},
  \citenamefont {Jung}, \citenamefont {Park}, \citenamefont {Yoon},
  \citenamefont {Lee}, \citenamefont {Kim}, \citenamefont {Choi}, \citenamefont
  {Takahashi},\ and\ \citenamefont {Hasko}}]{Shin2010}%
  \BibitemOpen
  \bibfield  {author} {\bibinfo {author} {\bibfnamefont {S.~J.}\ \bibnamefont
  {Shin}}, \bibinfo {author} {\bibfnamefont {C.~S.}\ \bibnamefont {Jung}},
  \bibinfo {author} {\bibfnamefont {B.~J.}\ \bibnamefont {Park}}, \bibinfo
  {author} {\bibfnamefont {T.~K.}\ \bibnamefont {Yoon}}, \bibinfo {author}
  {\bibfnamefont {J.~J.}\ \bibnamefont {Lee}}, \bibinfo {author} {\bibfnamefont
  {S.~J.}\ \bibnamefont {Kim}}, \bibinfo {author} {\bibfnamefont {J.~B.}\
  \bibnamefont {Choi}}, \bibinfo {author} {\bibfnamefont {Y.}~\bibnamefont
  {Takahashi}}, \ and\ \bibinfo {author} {\bibfnamefont {D.~G.}\ \bibnamefont
  {Hasko}},\ }\href {\doibase 10.1063/1.3483618} {\bibfield  {journal}
  {\bibinfo  {journal} {App. Phys. Lett.}\ }\textbf {\bibinfo {volume} {97}},\
  \bibinfo {pages} {103101} (\bibinfo {year} {2010})}\BibitemShut {NoStop}%
\bibitem [{\citenamefont {Knobel}\ and\ \citenamefont
  {Cleland}(2003)}]{Knobel2003}%
  \BibitemOpen
  \bibfield  {author} {\bibinfo {author} {\bibfnamefont {R.~G.}\ \bibnamefont
  {Knobel}}\ and\ \bibinfo {author} {\bibfnamefont {A.~N.}\ \bibnamefont
  {Cleland}},\ }\href@noop {} {\bibfield  {journal} {\bibinfo  {journal}
  {Nature}\ }\textbf {\bibinfo {volume} {424}},\ \bibinfo {pages} {291}
  (\bibinfo {year} {2003})}\BibitemShut {NoStop}%
\bibitem [{\citenamefont {Stokes}\ and\ \citenamefont
  {Khondaker}(2008)}]{Paul2008}%
  \BibitemOpen
  \bibfield  {author} {\bibinfo {author} {\bibfnamefont {P.}~\bibnamefont
  {Stokes}}\ and\ \bibinfo {author} {\bibfnamefont {S.~I.}\ \bibnamefont
  {Khondaker}},\ }\href@noop {} {\bibfield  {journal} {\bibinfo  {journal}
  {Appl. Phys. Lett.}\ }\textbf {\bibinfo {volume} {92}},\ \bibinfo {eid}
  {262107} (\bibinfo {year} {2008})}\BibitemShut {NoStop}%
\bibitem [{\citenamefont {Thelander}\ \emph {et~al.}(2001)\citenamefont
  {Thelander}, \citenamefont {Magnusson}, \citenamefont {Deppert},
  \citenamefont {Samuelson}, \citenamefont {Poulsen}, \citenamefont
  {Nygård},\ and\ \citenamefont {Borggreen}}]{Thelander2001}%
  \BibitemOpen
  \bibfield  {author} {\bibinfo {author} {\bibfnamefont {C.}~\bibnamefont
  {Thelander}}, \bibinfo {author} {\bibfnamefont {M.~H.}\ \bibnamefont
  {Magnusson}}, \bibinfo {author} {\bibfnamefont {K.}~\bibnamefont {Deppert}},
  \bibinfo {author} {\bibfnamefont {L.}~\bibnamefont {Samuelson}}, \bibinfo
  {author} {\bibfnamefont {P.~R.}\ \bibnamefont {Poulsen}}, \bibinfo {author}
  {\bibfnamefont {J.}~\bibnamefont {Nygård}}, \ and\ \bibinfo {author}
  {\bibfnamefont {J.}~\bibnamefont {Borggreen}},\ }\href {\doibase
  10.1063/1.1405154} {\bibfield  {journal} {\bibinfo  {journal} {Appl. Phys.
  Lett.}\ }\textbf {\bibinfo {volume} {79}},\ \bibinfo {pages} {2106} (\bibinfo
  {year} {2001})}\BibitemShut {NoStop}%
\bibitem [{\citenamefont {Klein}\ \emph {et~al.}(1997)\citenamefont {Klein},
  \citenamefont {Roth}, \citenamefont {Lim}, \citenamefont {Alivisatos},\ and\
  \citenamefont {McEuen}}]{Klein1997}%
  \BibitemOpen
  \bibfield  {author} {\bibinfo {author} {\bibfnamefont {D.~L.}\ \bibnamefont
  {Klein}}, \bibinfo {author} {\bibfnamefont {R.}~\bibnamefont {Roth}},
  \bibinfo {author} {\bibfnamefont {A.~K.~L.}\ \bibnamefont {Lim}}, \bibinfo
  {author} {\bibfnamefont {A.~P.}\ \bibnamefont {Alivisatos}}, \ and\ \bibinfo
  {author} {\bibfnamefont {P.~L.}\ \bibnamefont {McEuen}},\ }\href@noop {}
  {\bibfield  {journal} {\bibinfo  {journal} {Nature}\ }\textbf {\bibinfo
  {volume} {389}},\ \bibinfo {pages} {699} (\bibinfo {year}
  {1997})}\BibitemShut {NoStop}%
\bibitem [{\citenamefont {Kim}\ \emph {et~al.}(2002)\citenamefont {Kim},
  \citenamefont {Choo}, \citenamefont {Shim},\ and\ \citenamefont
  {Kang}}]{Kim2002}%
  \BibitemOpen
  \bibfield  {author} {\bibinfo {author} {\bibfnamefont {T.~W.}\ \bibnamefont
  {Kim}}, \bibinfo {author} {\bibfnamefont {D.~C.}\ \bibnamefont {Choo}},
  \bibinfo {author} {\bibfnamefont {J.~H.}\ \bibnamefont {Shim}}, \ and\
  \bibinfo {author} {\bibfnamefont {S.~O.}\ \bibnamefont {Kang}},\ }\href
  {\doibase 10.1063/1.1458685} {\bibfield  {journal} {\bibinfo  {journal}
  {Appl. Phys. Lett.}\ }\textbf {\bibinfo {volume} {80}},\ \bibinfo {pages}
  {2168} (\bibinfo {year} {2002})}\BibitemShut {NoStop}%
\bibitem [{\citenamefont {Stampfer}\ \emph {et~al.}(2008)\citenamefont
  {Stampfer}, \citenamefont {Schurtenberger}, \citenamefont {Molitor},
  \citenamefont {G\"{u}ttinger}, \citenamefont {Ihn},\ and\ \citenamefont
  {Ensslin}}]{Stampfer2008}%
  \BibitemOpen
  \bibfield  {author} {\bibinfo {author} {\bibfnamefont {C.}~\bibnamefont
  {Stampfer}}, \bibinfo {author} {\bibfnamefont {E.}~\bibnamefont
  {Schurtenberger}}, \bibinfo {author} {\bibfnamefont {F.}~\bibnamefont
  {Molitor}}, \bibinfo {author} {\bibfnamefont {J.}~\bibnamefont
  {G\"{u}ttinger}}, \bibinfo {author} {\bibfnamefont {T.}~\bibnamefont {Ihn}},
  \ and\ \bibinfo {author} {\bibfnamefont {K.}~\bibnamefont {Ensslin}},\ }\href
  {\doibase 10.1021/nl801225h} {\bibfield  {journal} {\bibinfo  {journal} {Nano
  Lett.}\ }\textbf {\bibinfo {volume} {8}},\ \bibinfo {pages} {2378} (\bibinfo
  {year} {2008})}\BibitemShut {NoStop}%
\bibitem [{\citenamefont {Song}\ \emph {et~al.}(2015)\citenamefont {Song},
  \citenamefont {Liu}, \citenamefont {Mosallanejad}, \citenamefont {You},
  \citenamefont {Han}, \citenamefont {Chen}, \citenamefont {Li}, \citenamefont
  {Cao}, \citenamefont {Xiao}, \citenamefont {Guo},\ and\ \citenamefont
  {Guo}}]{Song2015}%
  \BibitemOpen
  \bibfield  {author} {\bibinfo {author} {\bibfnamefont {X.-X.}\ \bibnamefont
  {Song}}, \bibinfo {author} {\bibfnamefont {D.}~\bibnamefont {Liu}}, \bibinfo
  {author} {\bibfnamefont {V.}~\bibnamefont {Mosallanejad}}, \bibinfo {author}
  {\bibfnamefont {J.}~\bibnamefont {You}}, \bibinfo {author} {\bibfnamefont
  {T.-Y.}\ \bibnamefont {Han}}, \bibinfo {author} {\bibfnamefont {D.-T.}\
  \bibnamefont {Chen}}, \bibinfo {author} {\bibfnamefont {H.-O.}\ \bibnamefont
  {Li}}, \bibinfo {author} {\bibfnamefont {G.}~\bibnamefont {Cao}}, \bibinfo
  {author} {\bibfnamefont {M.}~\bibnamefont {Xiao}}, \bibinfo {author}
  {\bibfnamefont {G.-C.}\ \bibnamefont {Guo}}, \ and\ \bibinfo {author}
  {\bibfnamefont {G.-P.}\ \bibnamefont {Guo}},\ }\href {\doibase
  10.1039/C5NR04961J} {\bibfield  {journal} {\bibinfo  {journal} {Nanoscale}\
  }\textbf {\bibinfo {volume} {7}},\ \bibinfo {pages} {16867} (\bibinfo {year}
  {2015})}\BibitemShut {NoStop}%
\bibitem [{\citenamefont {Angus}\ \emph {et~al.}(2007)\citenamefont {Angus},
  \citenamefont {Ferguson}, \citenamefont {Dzurak},\ and\ \citenamefont
  {Clark}}]{Angus2007}%
  \BibitemOpen
  \bibfield  {author} {\bibinfo {author} {\bibfnamefont {S.~J.}\ \bibnamefont
  {Angus}}, \bibinfo {author} {\bibfnamefont {A.~J.}\ \bibnamefont {Ferguson}},
  \bibinfo {author} {\bibfnamefont {A.~S.}\ \bibnamefont {Dzurak}}, \ and\
  \bibinfo {author} {\bibfnamefont {R.~G.}\ \bibnamefont {Clark}},\ }\href
  {\doibase 10.1021/nl070949k} {\bibfield  {journal} {\bibinfo  {journal} {Nano
  Lett.}\ }\textbf {\bibinfo {volume} {7}},\ \bibinfo {pages} {2051} (\bibinfo
  {year} {2007})}\BibitemShut {NoStop}%
\bibitem [{\citenamefont {Ramasubramaniam}\ \emph {et~al.}(2011)\citenamefont
  {Ramasubramaniam}, \citenamefont {Naveh},\ and\ \citenamefont
  {Towe}}]{PhysRevB}%
  \BibitemOpen
  \bibfield  {author} {\bibinfo {author} {\bibfnamefont {A.}~\bibnamefont
  {Ramasubramaniam}}, \bibinfo {author} {\bibfnamefont {D.}~\bibnamefont
  {Naveh}}, \ and\ \bibinfo {author} {\bibfnamefont {E.}~\bibnamefont {Towe}},\
  }\href {\doibase 10.1103/PhysRevB.84.205325} {\bibfield  {journal} {\bibinfo
  {journal} {Phys. Rev. B}\ }\textbf {\bibinfo {volume} {84}},\ \bibinfo
  {pages} {205325} (\bibinfo {year} {2011})}\BibitemShut {NoStop}%
\bibitem [{\citenamefont {Liu}\ \emph {et~al.}(2012)\citenamefont {Liu},
  \citenamefont {Li}, \citenamefont {Li}, \citenamefont {Gao}, \citenamefont
  {Chen},\ and\ \citenamefont {Lu}}]{Liu2012}%
  \BibitemOpen
  \bibfield  {author} {\bibinfo {author} {\bibfnamefont {Q.}~\bibnamefont
  {Liu}}, \bibinfo {author} {\bibfnamefont {L.}~\bibnamefont {Li}}, \bibinfo
  {author} {\bibfnamefont {Y.}~\bibnamefont {Li}}, \bibinfo {author}
  {\bibfnamefont {Z.}~\bibnamefont {Gao}}, \bibinfo {author} {\bibfnamefont
  {Z.}~\bibnamefont {Chen}}, \ and\ \bibinfo {author} {\bibfnamefont
  {J.}~\bibnamefont {Lu}},\ }\href {\doibase 10.1021/jp307124d} {\bibfield
  {journal} {\bibinfo  {journal} {J. Phys. Chem. C}\ }\textbf {\bibinfo
  {volume} {116}},\ \bibinfo {pages} {21556} (\bibinfo {year}
  {2012})}\BibitemShut {NoStop}%
\bibitem [{com()}]{comsol}%
  \BibitemOpen
  \href@noop {} {}\bibinfo {note} {\textsc{comsol
  multiphysics}\textsuperscript{\textregistered} version 5.1, COMSOL, Inc.,
  Burlington, MA, USA, 2008.}\BibitemShut {Stop}%
\bibitem [{\citenamefont {Dovesi}\ \emph {et~al.}(2005)\citenamefont {Dovesi},
  \citenamefont {Orlando}, \citenamefont {Civalleri}, \citenamefont {Roetti},
  \citenamefont {Saunders},\ and\ \citenamefont
  {Zicovich-Wilson}}]{dovesi2005crystal}%
  \BibitemOpen
  \bibfield  {author} {\bibinfo {author} {\bibfnamefont {R.}~\bibnamefont
  {Dovesi}}, \bibinfo {author} {\bibfnamefont {R.}~\bibnamefont {Orlando}},
  \bibinfo {author} {\bibfnamefont {B.}~\bibnamefont {Civalleri}}, \bibinfo
  {author} {\bibfnamefont {C.}~\bibnamefont {Roetti}}, \bibinfo {author}
  {\bibfnamefont {V.~R.}\ \bibnamefont {Saunders}}, \ and\ \bibinfo {author}
  {\bibfnamefont {C.~M.}\ \bibnamefont {Zicovich-Wilson}},\ }\href@noop {}
  {\bibfield  {journal} {\bibinfo  {journal} {Zeitschrift f{\"u}r
  Kristallographie}\ }\textbf {\bibinfo {volume} {220}},\ \bibinfo {pages}
  {571} (\bibinfo {year} {2005})}\BibitemShut {NoStop}%
\bibitem [{\citenamefont {Dovesi}\ \emph {et~al.}(2013)\citenamefont {Dovesi},
  \citenamefont {Saunders}, \citenamefont {Roetti}, \citenamefont {Orlando},
  \citenamefont {Pascale}, \citenamefont {Civalleri}, \citenamefont {Doll},
  \citenamefont {Harrison}, \citenamefont {Bush}, \citenamefont {Llunell},
  \citenamefont {Science},\ and\ \citenamefont {Technologies}}]{Dovesi2013}%
  \BibitemOpen
  \bibfield  {author} {\bibinfo {author} {\bibfnamefont {R.}~\bibnamefont
  {Dovesi}}, \bibinfo {author} {\bibfnamefont {V.~R.}\ \bibnamefont
  {Saunders}}, \bibinfo {author} {\bibfnamefont {C.}~\bibnamefont {Roetti}},
  \bibinfo {author} {\bibfnamefont {R.}~\bibnamefont {Orlando}}, \bibinfo
  {author} {\bibfnamefont {F.}~\bibnamefont {Pascale}}, \bibinfo {author}
  {\bibfnamefont {B.}~\bibnamefont {Civalleri}}, \bibinfo {author}
  {\bibfnamefont {K.}~\bibnamefont {Doll}}, \bibinfo {author} {\bibfnamefont
  {N.~M.}\ \bibnamefont {Harrison}}, \bibinfo {author} {\bibfnamefont {I.~J.}\
  \bibnamefont {Bush}}, \bibinfo {author} {\bibfnamefont {M.}~\bibnamefont
  {Llunell}}, \bibinfo {author} {\bibfnamefont {C.}~\bibnamefont {Science}}, \
  and\ \bibinfo {author} {\bibfnamefont {A.}~\bibnamefont {Technologies}},\
  }\href {\doibase 10.1524/zkri.220.5.571.65065} {\emph {\bibinfo {title}
  {{Crystal09 User's Manual}}}} (\bibinfo {year} {2013})\BibitemShut {NoStop}%
\bibitem [{\citenamefont {Dickinson}\ and\ \citenamefont
  {Pauling}(1923)}]{Dickinson1923}%
  \BibitemOpen
  \bibfield  {author} {\bibinfo {author} {\bibfnamefont {R.~G.}\ \bibnamefont
  {Dickinson}}\ and\ \bibinfo {author} {\bibfnamefont {L.}~\bibnamefont
  {Pauling}},\ }\href {\doibase 10.1021/ja01659a020} {\bibfield  {journal}
  {\bibinfo  {journal} {J. Am. Chem. Soc.}\ }\textbf {\bibinfo {volume} {45}},\
  \bibinfo {pages} {1466} (\bibinfo {year} {1923})}\BibitemShut {NoStop}%
\bibitem [{\citenamefont {Perdew}\ \emph {et~al.}(2008)\citenamefont {Perdew},
  \citenamefont {Ruzsinszky}, \citenamefont {Csonka}, \citenamefont {Vydrov},
  \citenamefont {Scuseria}, \citenamefont {Constantin}, \citenamefont {Zhou},\
  and\ \citenamefont {Burke}}]{Perdew2008}%
  \BibitemOpen
  \bibfield  {author} {\bibinfo {author} {\bibfnamefont {J.~P.}\ \bibnamefont
  {Perdew}}, \bibinfo {author} {\bibfnamefont {A.}~\bibnamefont {Ruzsinszky}},
  \bibinfo {author} {\bibfnamefont {G.~I.}\ \bibnamefont {Csonka}}, \bibinfo
  {author} {\bibfnamefont {O.~A.}\ \bibnamefont {Vydrov}}, \bibinfo {author}
  {\bibfnamefont {G.~E.}\ \bibnamefont {Scuseria}}, \bibinfo {author}
  {\bibfnamefont {L.~A.}\ \bibnamefont {Constantin}}, \bibinfo {author}
  {\bibfnamefont {X.}~\bibnamefont {Zhou}}, \ and\ \bibinfo {author}
  {\bibfnamefont {K.}~\bibnamefont {Burke}},\ }\href {\doibase
  10.1103/PhysRevLett.100.136406} {\bibfield  {journal} {\bibinfo  {journal}
  {Phys. Rev. Lett.}\ }\textbf {\bibinfo {volume} {100}},\ \bibinfo {pages}
  {136406} (\bibinfo {year} {2008})}\BibitemShut {NoStop}%
\bibitem [{\citenamefont {Cor{\`a}}\ \emph {et~al.}(1997)\citenamefont
  {Cor{\`a}}, \citenamefont {Patel}, \citenamefont {Harrison}, \citenamefont
  {Roetti},\ and\ \citenamefont {Catlow}}]{cora1997anab}%
  \BibitemOpen
  \bibfield  {author} {\bibinfo {author} {\bibfnamefont {F.}~\bibnamefont
  {Cor{\`a}}}, \bibinfo {author} {\bibfnamefont {A.}~\bibnamefont {Patel}},
  \bibinfo {author} {\bibfnamefont {N.~M.}\ \bibnamefont {Harrison}}, \bibinfo
  {author} {\bibfnamefont {C.}~\bibnamefont {Roetti}}, \ and\ \bibinfo {author}
  {\bibfnamefont {C.~R.~A.}\ \bibnamefont {Catlow}},\ }\href@noop {} {\bibfield
   {journal} {\bibinfo  {journal} {J. Mater. Chem.}\ }\textbf {\bibinfo
  {volume} {7}},\ \bibinfo {pages} {959} (\bibinfo {year} {1997})}\BibitemShut
  {NoStop}%
\bibitem [{\citenamefont {Ehlers}\ \emph {et~al.}(1993)\citenamefont {Ehlers},
  \citenamefont {B{\"{o}}hme}, \citenamefont {Dapprich}, \citenamefont {Gobbi},
  \citenamefont {H{\"{o}}llwarth}, \citenamefont {Jonas}, \citenamefont
  {K{\"{o}}hler}, \citenamefont {Stegmann}, \citenamefont {Veldkamp},\ and\
  \citenamefont {Frenking}}]{Ehlers1993}%
  \BibitemOpen
  \bibfield  {author} {\bibinfo {author} {\bibfnamefont {A.}~\bibnamefont
  {Ehlers}}, \bibinfo {author} {\bibfnamefont {M.}~\bibnamefont {B{\"{o}}hme}},
  \bibinfo {author} {\bibfnamefont {S.}~\bibnamefont {Dapprich}}, \bibinfo
  {author} {\bibfnamefont {A.}~\bibnamefont {Gobbi}}, \bibinfo {author}
  {\bibfnamefont {A.}~\bibnamefont {H{\"{o}}llwarth}}, \bibinfo {author}
  {\bibfnamefont {V.}~\bibnamefont {Jonas}}, \bibinfo {author} {\bibfnamefont
  {K.}~\bibnamefont {K{\"{o}}hler}}, \bibinfo {author} {\bibfnamefont
  {R.}~\bibnamefont {Stegmann}}, \bibinfo {author} {\bibfnamefont
  {A.}~\bibnamefont {Veldkamp}}, \ and\ \bibinfo {author} {\bibfnamefont
  {G.}~\bibnamefont {Frenking}},\ }\href {\doibase
  10.1016/0009-2614(93)80086-5} {\bibfield  {journal} {\bibinfo  {journal}
  {Chem. Phys. Lett.}\ }\textbf {\bibinfo {volume} {208}},\ \bibinfo {pages}
  {111} (\bibinfo {year} {1993})}\BibitemShut {NoStop}%
\bibitem [{\citenamefont {Peintinger}\ \emph {et~al.}(2013)\citenamefont
  {Peintinger}, \citenamefont {Oliveira},\ and\ \citenamefont
  {Bredow}}]{Peintinger2013}%
  \BibitemOpen
  \bibfield  {author} {\bibinfo {author} {\bibfnamefont {M.~F.}\ \bibnamefont
  {Peintinger}}, \bibinfo {author} {\bibfnamefont {D.~V.}\ \bibnamefont
  {Oliveira}}, \ and\ \bibinfo {author} {\bibfnamefont {T.}~\bibnamefont
  {Bredow}},\ }\href {\doibase 10.1002/jcc.23153} {\bibfield  {journal}
  {\bibinfo  {journal} {J. Comput. Chem.}\ }\textbf {\bibinfo {volume} {34}},\
  \bibinfo {pages} {451} (\bibinfo {year} {2013})}\BibitemShut {NoStop}%
\bibitem [{\citenamefont {Monkhorst}\ and\ \citenamefont
  {Pack}(1976)}]{Monkhorst1976}%
  \BibitemOpen
  \bibfield  {author} {\bibinfo {author} {\bibfnamefont {H.~J.}\ \bibnamefont
  {Monkhorst}}\ and\ \bibinfo {author} {\bibfnamefont {J.~D.}\ \bibnamefont
  {Pack}},\ }\href {\doibase 10.1103/PhysRevB.13.5188} {\bibfield  {journal}
  {\bibinfo  {journal} {Phys. Rev. B}\ }\textbf {\bibinfo {volume} {13}},\
  \bibinfo {pages} {5188} (\bibinfo {year} {1976})}\BibitemShut {NoStop}%
\bibitem [{\citenamefont {van~der Wiel}\ \emph {et~al.}(2002)\citenamefont
  {van~der Wiel}, \citenamefont {De~Franceschi}, \citenamefont {Elzerman},
  \citenamefont {Fujisawa}, \citenamefont {Tarucha},\ and\ \citenamefont
  {Kouwenhoven}}]{RevModPhys.75.1}%
  \BibitemOpen
  \bibfield  {author} {\bibinfo {author} {\bibfnamefont {W.~G.}\ \bibnamefont
  {van~der Wiel}}, \bibinfo {author} {\bibfnamefont {S.}~\bibnamefont
  {De~Franceschi}}, \bibinfo {author} {\bibfnamefont {J.~M.}\ \bibnamefont
  {Elzerman}}, \bibinfo {author} {\bibfnamefont {T.}~\bibnamefont {Fujisawa}},
  \bibinfo {author} {\bibfnamefont {S.}~\bibnamefont {Tarucha}}, \ and\
  \bibinfo {author} {\bibfnamefont {L.~P.}\ \bibnamefont {Kouwenhoven}},\
  }\href {\doibase 10.1103/RevModPhys.75.1} {\bibfield  {journal} {\bibinfo
  {journal} {Rev. Mod. Phys.}\ }\textbf {\bibinfo {volume} {75}},\ \bibinfo
  {pages} {1} (\bibinfo {year} {2002})}\BibitemShut {NoStop}%
\bibitem [{\citenamefont {Jones}\ \emph {et~al.}(2005)\citenamefont {Jones},
  \citenamefont {Kwon},\ and\ \citenamefont {Norton}}]{Jones2005}%
  \BibitemOpen
  \bibfield  {author} {\bibinfo {author} {\bibfnamefont {M.}~\bibnamefont
  {Jones}}, \bibinfo {author} {\bibfnamefont {Y.}~\bibnamefont {Kwon}}, \ and\
  \bibinfo {author} {\bibfnamefont {D.}~\bibnamefont {Norton}},\ }\href
  {\doibase 10.1007/s00339-005-3208-2} {\bibfield  {journal} {\bibinfo
  {journal} {Appl. Phys. A}\ }\textbf {\bibinfo {volume} {81}},\ \bibinfo
  {pages} {285} (\bibinfo {year} {2005})}\BibitemShut {NoStop}%
\bibitem [{\citenamefont {Cheiwchanchamnangij}\ and\ \citenamefont
  {Lambrecht}(2012)}]{mass}%
  \BibitemOpen
  \bibfield  {author} {\bibinfo {author} {\bibfnamefont {T.}~\bibnamefont
  {Cheiwchanchamnangij}}\ and\ \bibinfo {author} {\bibfnamefont {W.~R.~L.}\
  \bibnamefont {Lambrecht}},\ }\href {\doibase 10.1103/PhysRevB.85.205302}
  {\bibfield  {journal} {\bibinfo  {journal} {Phys. Rev. B}\ }\textbf {\bibinfo
  {volume} {85}},\ \bibinfo {pages} {205302} (\bibinfo {year}
  {2012})}\BibitemShut {NoStop}%
\bibitem [{\citenamefont {Koppinen}(2003)}]{thesis}%
  \BibitemOpen
  \bibfield  {author} {\bibinfo {author} {\bibfnamefont {P.}~\bibnamefont
  {Koppinen}},\ }\emph {\bibinfo {title} {Bias and temperature dependence
  analysis of the tunneling current of normal metal-insulator-normal metal
  tunnel junctions}},\ \href@noop {} {Master's thesis},\ \bibinfo  {school}
  {University of Jyvaskyla, department of physics} (\bibinfo {year}
  {2003})\BibitemShut {NoStop}%
\bibitem [{\citenamefont {Conrad}\ \emph {et~al.}(2005)\citenamefont {Conrad},
  \citenamefont {Greentree}, \citenamefont {Jamieson},\ and\ \citenamefont
  {Hollenberg}}]{Conrad05}%
  \BibitemOpen
  \bibfield  {author} {\bibinfo {author} {\bibfnamefont {V.~I.}\ \bibnamefont
  {Conrad}}, \bibinfo {author} {\bibfnamefont {A.~D.}\ \bibnamefont
  {Greentree}}, \bibinfo {author} {\bibfnamefont {D.~N.}\ \bibnamefont
  {Jamieson}}, \ and\ \bibinfo {author} {\bibfnamefont {L.~C.~L.}\ \bibnamefont
  {Hollenberg}},\ }\href {\doibase doi:10.1166/jctn.2005.105} {\bibfield
  {journal} {\bibinfo  {journal} {J. Comput. Theor. Nanos.}\ }\textbf {\bibinfo
  {volume} {2}},\ \bibinfo {pages} {214} (\bibinfo {year} {2005})}\BibitemShut
  {NoStop}%
\end{thebibliography}%

\end{document}